\documentclass[preprint,12pt]{elsarticle}

\usepackage{graphicx}
\usepackage{amssymb}

\usepackage{lineno}

\journal{Journal Name}

\begin{document}

\begin{frontmatter}


\title{OAM beams from incomplete computer generated holograms}



\author[1]{Ni\~na Angelica F. Zambale}
\author[1]{Gerald John H. Doblado}
\author[1]{Nathaniel Hermosa}

\address{National Institute of Physics, University of the Philippines Diliman, Quezon City 1101 Philippines}

\begin{abstract}
We show that optical beams with orbital angular momentum (OAM) can be generated even with incomplete computer generated holograms (CGH). These holograms are made such that random portions of it do not contain any information. We observe that although the beams produced with these holograms are less intense, these beams maintain their shape and that their topological charges are not affected. Furthermore, we show that superposition of two or more beams can be created using separate incomplete CGHs interspersed together. Our result is significant especially since most method to generate beams with OAM for various applications rely on pixelated devices or optical elements with imperfections.
\end{abstract}


\end{frontmatter}


\section{Introduction}
\label{S:1}

The realization that light beams can have quantized orbital angular momentum in addition to spin angular momentum has led, in recent years, to novel experiments in quantum and classical optics\cite{padgett2004light, allen1999iv,mair2001entanglement, MeranoPRA2010}, new methods for manipulating micro particles \cite{galajda2001complex, gahagan1996optical}, new possibilities in optical metrology \cite{hell1994breaking, rosales2013experimental} and new ways to boost the capacity of communication channel \cite{bozinovic2013terabit} to name a few of the many exciting applications of these beams. Optical beams with OAM are best prepared with helically phased light beams, such as the Laguerre Gaussian (LG) modes, which have an explicit $\ell\phi$ phase factor, where $\ell$ is the topological charge and $\phi$ is the azimuthal coordinate \cite{allen1992orbital, van1992eigenfunction}. This phase factor makes them natural choice for describing beams carrying orbital angular momentum \cite{padgett2004light,allen1992orbital, van1992eigenfunction}.

There are a number of ways to generate beams with OAM. These beams can be created either with passive optical elements and devices or with nonlinear optical materials \cite{allen1999iv}. Passive optical elements and devices include pitched-forked hologram\cite{bazhenov1990laser,heckenberg1992laser, heckenberg1992generation}, astigmatic mode converter \cite{allen1992orbital, beijersbergen1993astigmatic}, spiral phase plates \cite{beijersbergen1994helical}, the q-plate \cite{marrucci2006optical, karimi2009efficient}, achromatic OAM generators \cite{bouchard2014achromatic}, mirror cones \cite{mansuripur2011spin, kobayashi2012helical}, and Pancharatnam-Berry Phase optical elements \cite{marrucci2006pancharatnam, biener2002formation}.  OAM beams can also be produced with nonlinear crystals thru second harmonic generation \cite{dholakia1996second, zhou2014highly},  parametric down conversion \cite{mair2001entanglement}, sum-frequency generation\cite{zhou2014generation, steinlechner2016frequency} and echo-enabled harmonic generation \cite{hemsing2012echo}. Of these methods, the most widely used is the pitched-forked hologram where the pattern is either printed \cite{bazhenov1990laser}, mapped on to a device whose phase can be controlled such as an spatial light modulator (for example \cite{gibson2004free}) or programmed to a digital micromirror device (DMD) (such as in \cite{mirhosseini2013rapid}). Pitched-forked hologram are also used to generate electron beams with OAM \cite{verbeeck2010production}.

\begin{figure}[tb]
\centering \includegraphics[width=0.6\linewidth]{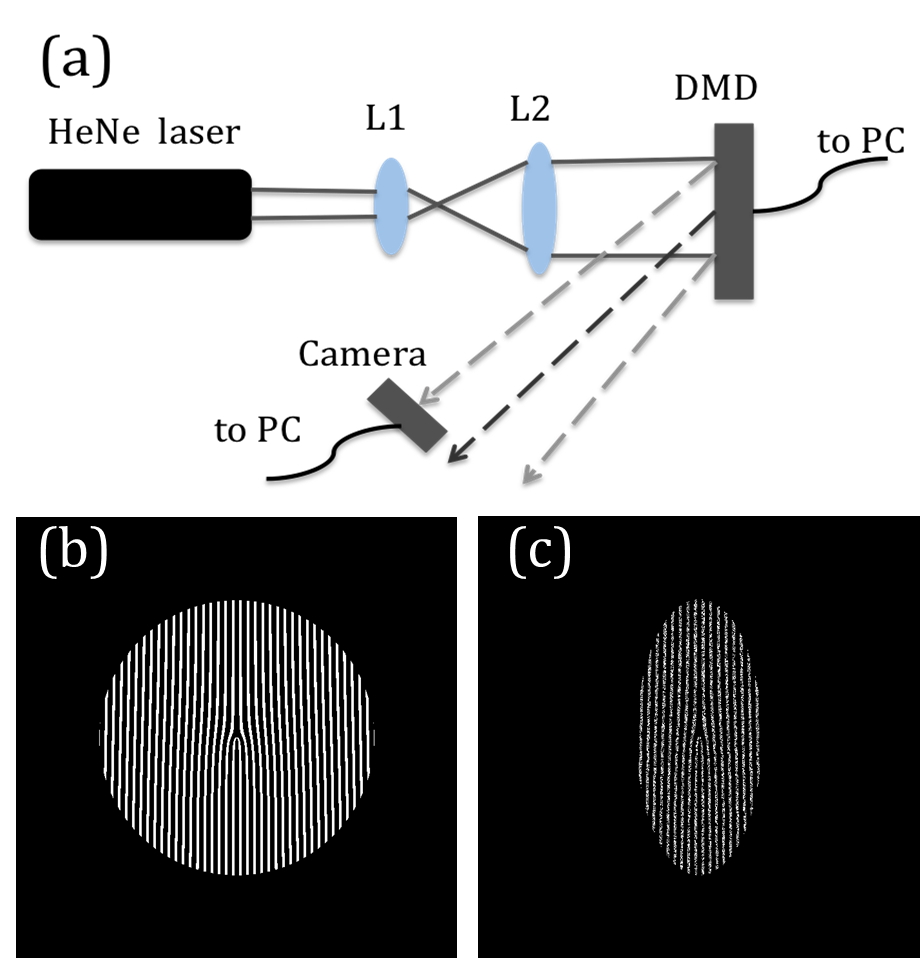}
\caption{(\emph{false color, color online}) (a) A standard optical setup to produce OAM beams with computer generated holograms (CGH). A HeNe laser ($\lambda$ = 632.8 nm ) is collimated, expanded and is incident onto a DMD. The first order diffraction from the DMD is captured with a camera. Sample CGHs are shown in  (b) for the original CGH and in (c) re-scaled CGH with removed pixels. The CGH is elongated in one axis to counter the inherent elongation being made by the DMD used.}
\label{fig:setup}
\end{figure}

In this paper, we present the use of incomplete forked holograms to produce beams with OAM.  The holograms are made such that in random portions of it, the information about the beam to be reconstructed are removed. Such a hologram happens in reality, most especially when the device is made up of an array of much smaller devices whose phase can be controlled as in the case of the SLM or whose side of beam's deflection can be programmed such as in DMD. Any of these tiny devices at some time may not work and hence, may have a considerable effect on the output beam. On the other hand, if by removing certain portions of the CGH will not have a significant effect, then it may be possible to put information from a hologram of another beam in these portions. The reconstructed beam is then a superposition of the beams produced by the  different holograms.  In effect, information can be placed in the device independently. These are what we intend to answer in this paper.

Intuitively, we can argue that  since beams with OAM self-reconstruct due to the beams' natural internal energy flow \cite{allen1992orbital,vaity2011self, broky2008self,bouchal2002resistance}, puncturing the CGHs that produce these beams would only have an effect on the intensity of the beam and that its topological charge will be conserved. Several researchers have performed experiments on beams with OAM and observed that when a certain part of the beam is blocked, the beam will be able to reconstruct at distances of the order of the Rayleigh length in the case of LG beams\cite{vaity2011self} or lesser in the case of higher-order Bessel beams and Helico-conical optical beams \cite{garces2002simultaneous,hermosa2013helico}. Our experiments here are different in that instead of blocking portions, we removed any information of the beam in the CGH randomly. Moreover, we replaced in those regions information from another CGH of a different beam. 

Similar works have also been done in the past regarding incomplete or interspersed holograms\cite{som1971new,caulfield1970wavefront}. Both papers discussed holographic multiplexing in film holograms wherein various masks were made to store information into a single film hologram. These masks were again used in the reconstruction to retain the desired image and remove all unwanted information.Researchers have found out that removing or masking portions of the holograms results into a decrease in intensity of the reconstructed image of the object proportional to $1/N^2$ where $N$ is the number of object wavefronts. It is generally accepted that in this setup, each area of the film hologram contains all the information about the object wavefront. Removing portions of it will only result in a decrease in intensity of the reconstruction instead of a loss of information\cite{som1971new}. However, the main difference between the works of Caulfield and Som and our work is that we use beams with orbital angular momentum as our object wavefront. The orbital angular momentum of the beam is embedded only on some areas of the hologram. If the portions that were randomly removed contains information regarding the orbital angular momentum of the beam, it is possible that the OAM of the reconstructed beam can never be recovered. 

Here, we provide experimental evidence that OAM beams can still be produced even with limited information in the CGH. Although the intensity is affected, the topological charge is not. We observe the intensity profile and we quantify the reduction of the intensity. We get interference patterns to prove that the topological charge does not change even at minimal information. Finally, we show that superposed beams can be produced from different holograms interspersed together. We have put together up to three holograms as proof of concept demonstration. 

\section{Methodology}
\label{S:2}

We generate beams with OAM using a DMD in a DLP Lightcrafter (Texas Instruments). With the light engine removed, we expose the array of micromirrors which are programmed to have a binary pitched-forked hologram. Figure \ref{fig:setup}(a) shows the experimental setup. A collimated HeNe laser ($\lambda$=632.8 nm) beam impinges onto the exposed micromirror arrays of the DMD. The first-order diffraction from the hologram is isolated and its intensity profile at a distance of 0.50 m from the DMD is captured by a CMOS camera (DMM 72BUC02, The Imaging Source). 

The DMD was encoded with binarized computer generated holograms. A beam $E_1$ with a phase of $\ell \phi$ was superposed to a reference beam with a carrier frequency to produce a hologram ($H$). Mathematically, the hologram is described by Eqn.\ref{eqn:hol} 

\begin{equation}
H = {|E_{ref} + E_1|}^2 = {|e^{i\vec{k} \cdot \vec{r}} + e^{i\ell\phi}|}^2
\label{eqn:hol}
\end{equation}

\begin{figure}[tb]
\centering \includegraphics[width=0.8\linewidth]{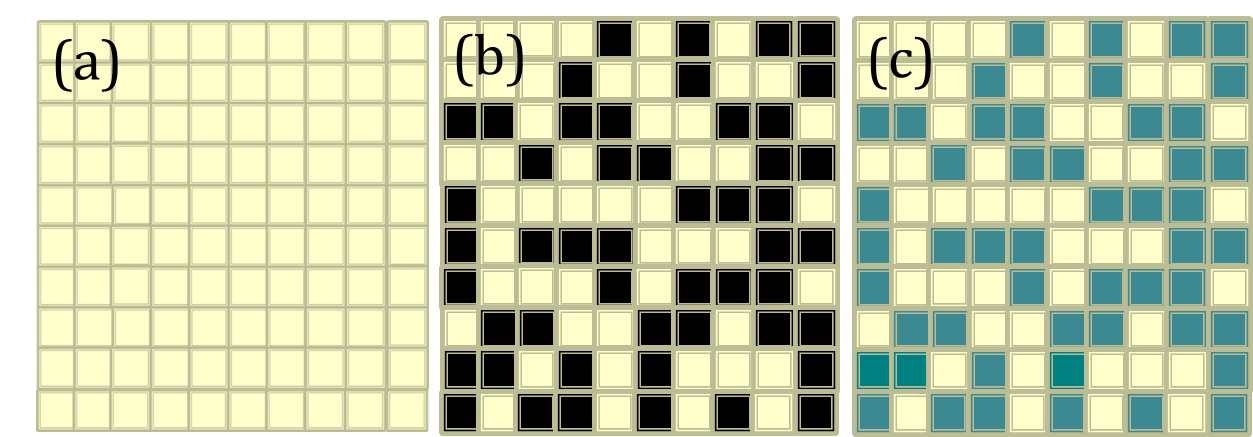}
\caption{(\emph{false color, color online}) Visual representation of the CGHs. (a) The complete CGH as a 10 $\times$ 10 grid and (b) the incomplete CGH with 50\% information retained where the black squares represent regions that have no information. (c) The superposition of two CGHs using the technique introduced. Each of the yellow and blue squares have information on two different CGHs. }
\label{fig:visual}
\end{figure}

\begin{figure}[t]
\centering \includegraphics[width=0.9\linewidth]{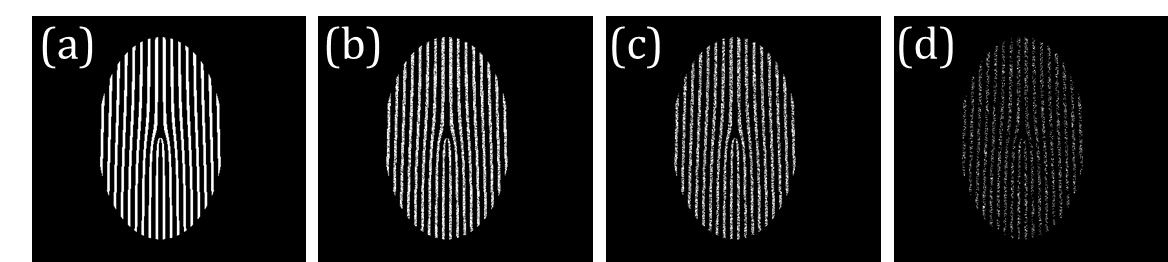}
\vspace{-1em}
\caption{(\emph{false color, color online}) The computer generated holograms based on the amount of information retained on the holograms.}
\label{fig:CGHs}
\end{figure}

A binarization scheme was implemented to the CGHs such that the gray values less than half the maximum become 0's and all the remaining pixels become 1's. A binary random mask with values 0 and 1 was then multiplied to the CGH to produce random portions that have no information. The randomization of the pixel values of the mask followed the random permutation function in MATLAB. Different CGHs were created for different values of topological charge $\ell$ and different binary masks were produced corresponding to different fractions of information that were removed. To further visualize the process of the removing information from the hologram, we make an analogy of our hologram and a 10 $\times$ 10 grid. Our complete hologram is shown in Fig.\ref{fig:visual}(a) where each pixel contains information. The removal of information is shown in Fig.\ref{fig:visual}(b) where 50\% of information was randomly removed. In this figure, the black portions of the grid represent the regions where there is zero information. These regions do not contribute to the reconstructed beam. Fig. \ref{fig:CGHs} shows the corresponding incomplete CGHs based on the amount of information retained on the holograms. 

We noticed that the DMD we used elongates the hologram in the horizontal axis, hence we calibrated our CGH accordingly such that at the DMD the vertical and the horizontal axis scales the same. The scaling factor $k$ was determined by the comparing the line scans of a simulated and experimental OAM beams. We present a sample of the original and re-scaled CGHs in fig.\ref{fig:setup}(b) and (c). 

The integrated optical densities (IOP) of the captured images from the incomplete holograms were compared to the IOP from a complete hologram. The IOP is effectively the total intensity of the beam. The IOP was calculated as the sum of gray values of an 8-bit image for all the pixels of the image, IOP = $\sum_i^N\sum_j^N gv_{ij}$ where $N \times N$ is the number of pixels and $gv_{ij}$ is the gray level of pixel at column $i$ and row $j$ .  We made sure that the camera captures all images at the same setting. 

We determined the topological charge of the generated beam from different CGHs by interfering it with a tilted plane wave. There are several ways to measure the topological charge of a beam. Some of the methods include the use of tilted spherical lens\cite{vaity2013measuring}, various apertures \cite{de2011measuring,guo2009characterizing} and interferometric method \cite{arlt1998production}. However, the interferometric method was chosen since the number of bifurcation in the interference pattern gives the topological charge of the beam automatically. We counted the number of bifurcation to determine the topological charge.

As proof-of-principle that superposed beam can be obtained from interspersed hologram, we replaced the removed portions with information from another CGH (Fig. \ref{fig:visual}(c)). In effect, we created a hologram of the form,

\begin{equation}
H_{combined}\sim H_1^i+H_2^{ii}+...H_n^{i^n}
\label{eqn:combined}
\end{equation} 

\noindent where $n$ is the $nth$ hologram of the beam that we want to reconstruct and the $i$'s indicate that these holograms are not added but their components are placed in different positions in the combined CGH, $H_{combined}$. We compare the intensity patterns we obtained from eqn.\ref{eqn:combined} with the generated beams from a CGH that was calculated from, 

\begin{equation}
H\sim|E_{ref}+E_1+E_2+...E_n|^2
\label{eqn:usual}
\end{equation}

\begin{figure}[tb]
\centering \includegraphics[width=0.65\linewidth]{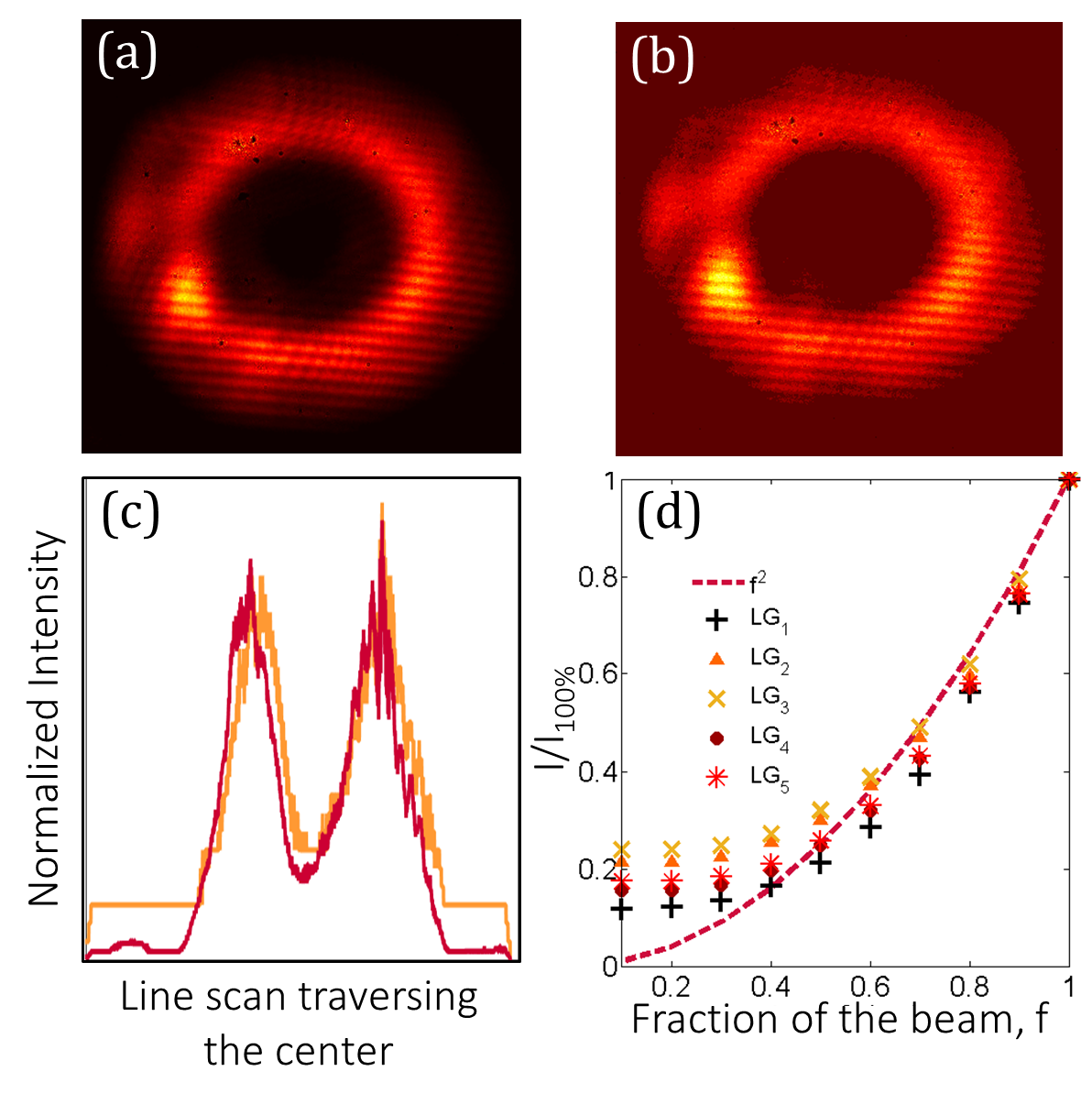}
\caption{(\emph{false color, color online}) Representative generated beams. (a) Generated beam with complete CGH. (b) Generated beam with randomly removed 50\% of information. (c) Linescan comparison between beams (a) in red and (b) in orange. (d) The integrated optical density (IOP) of the beams from incomplete CGHs with respect to the IOP of the complete CGH for different values of $\ell$ ($I/I_{100\%}$). $I/I_{100\%}$ follows a quadratic trend in $f$.}
\label{fig:profiles}
\end{figure}

\noindent where $H$ is the hologram, $n$ is the \emph{nth} beam to be added and $E_{ref}$ is the reference beam that is tilted to effect a carrier frequency. Eqn. \ref{eqn:usual} is of course the usual method of creating a CGH of superposed beams. By placing known beams' electric field expressions in $E$, a CGH of a superposition of these beams upon reconstruction will be produced.

Eqn. \ref{eqn:combined} is not equivalent to eqn. \ref{eqn:usual}. The hologram in eqn. \ref{eqn:usual} takes into account the interferences of the $E$'s while in eqn. \ref{eqn:combined} the $E$'s are independent of each other: the $E$'s interfering only with the $E_{ref}$ since $H_n\sim |E_{ref}+E_n|^2$. The $H_n$'s in eqn. \ref{eqn:combined} are related only because the placement of the elements of one hologram will depend on the placement of the elements of the other holograms. However, here we show that the resulting superposed beams are similar.

\section{Results and Discussion}

Figure \ref{fig:profiles} shows a sample of experimentally generated OAM beam. We show the intensity profiles for $\ell =2$ with the complete hologram (a) and with 50\% of information removed randomly. We have done the experiments for several other values of $\ell$'s and in general, we obtain the same results.The horizontal fringes present in the intensity profiles are artifacts of the optical elements used in the experiment. These fringes exist in all the intensity profiles, hence it does not affect the comparison between the reconstructed OAM beams. 

In \ref{fig:profiles}(c), we normalized the linescans to prove that indeed the shape is preserved.  The linescan for 50\% had a small intensity offset. Most beams reconstructed with incomplete holograms would have this offset especially when the fraction of the hologram removed is high. All $\ell$'s follow the same trend. The intensity of the beam is minutely affected by the $\ell$'s only and it is only visible at the extreme fraction of loss. This is important because by changing the fraction of removed information, one can quantify the amount of the beam that will be reconstructed. We can attenuate the beam's intensity at a controlled amount without even using a filter. This decrease in intensity with the amount of removed information is not surprising. A fraction $f$ removed from $H$ corresponds to an $f^2$ decrease in the intensity of the reconstructed beam. This is similar to the observation of Som in multispersed film holograms \cite{som1971new}. The additional offset in the reconstructed beam's intensity may be attributed to the randomly scattered beam by the graininess of the hologram.

\begin{figure}[t]
\centering \includegraphics[width=0.9\linewidth]{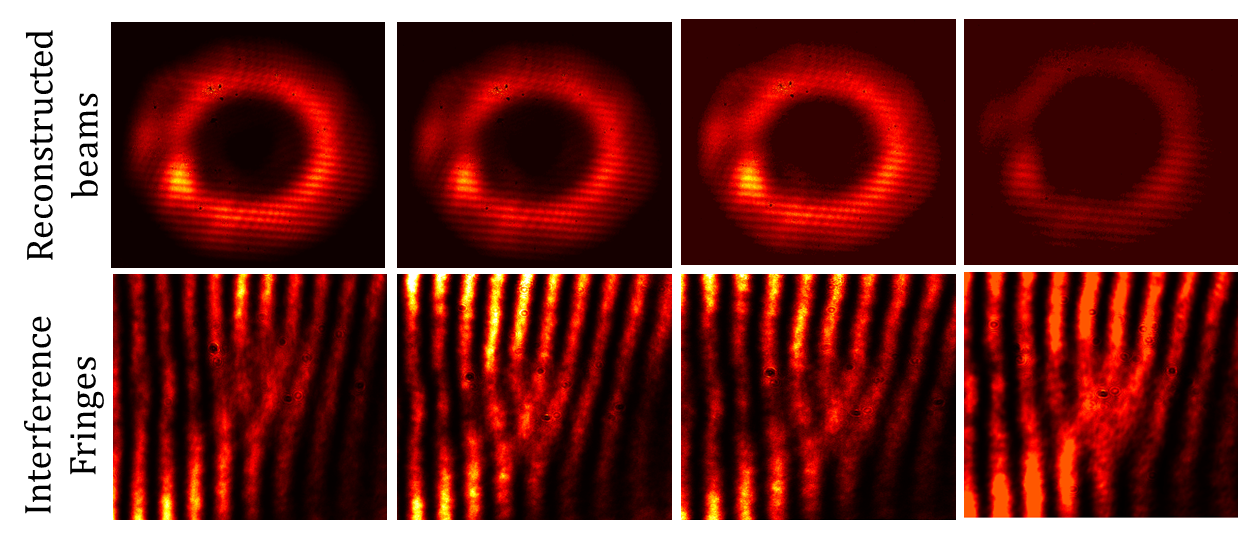}
\caption{(\emph{false color, color online}) Interference pattern of the beams ($\ell=2$ and with different completeness in \%) with a tilted plane wave showing the conservation of the topological charge even when the fraction of removed information is increased.}
\label{fig:topological}
\end{figure}

The topological charge of the beams can be obtained by interfering a tilted plane wave with the generated beams and looking at the resulting interference pattern. We show our results in fig. \ref{fig:topological} for $\ell=2$, although as we discussed earlier, the intensity is lessened. We observe that the topological charge is still conserved. This is an important observation of our work. Even though we have used incomplete computer generated holograms, the topological charge, hence the orbital angular momentum remains the same.

\begin{figure}[b!]
\centering \includegraphics[width=0.9\linewidth]{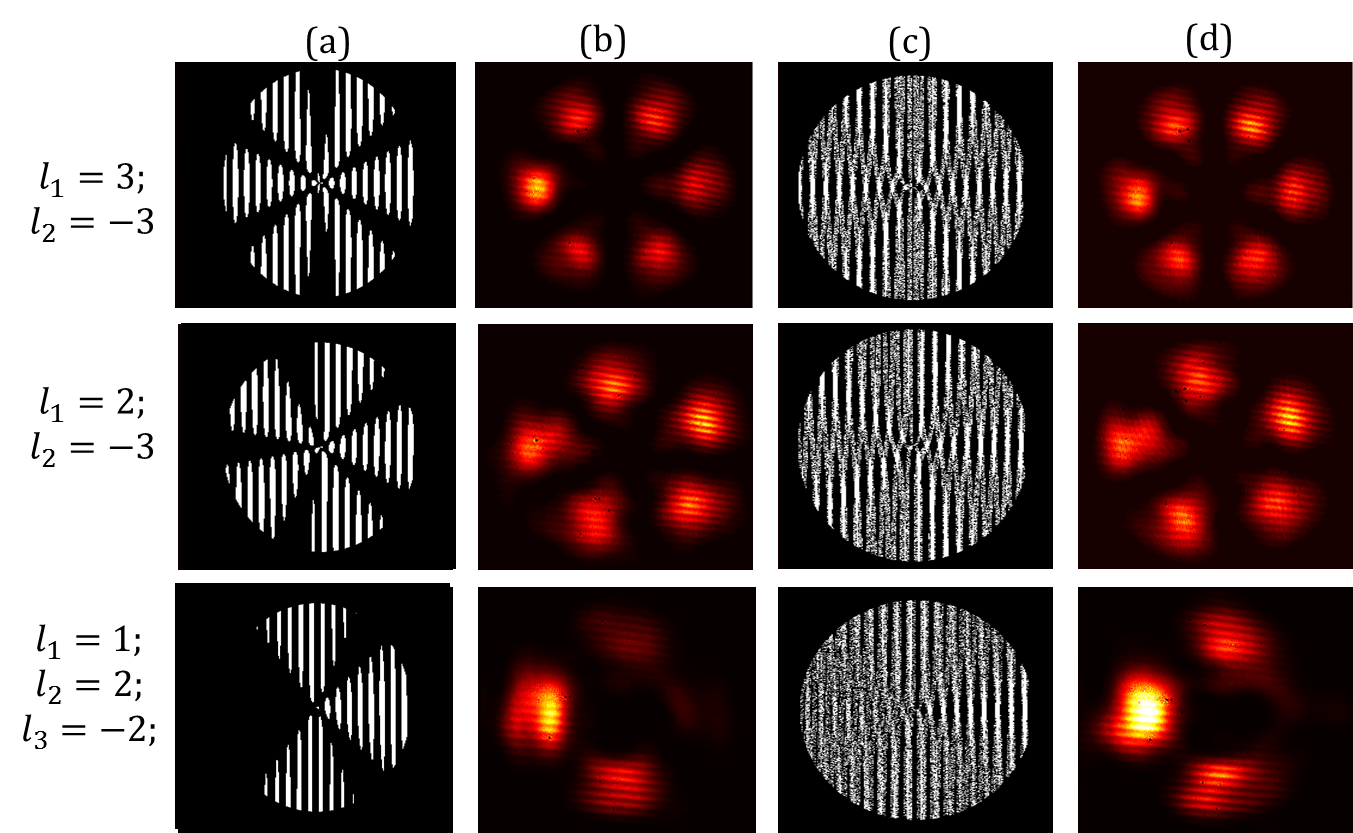}
\caption{(\emph{false color, color online}) Reconstructed superposed beam obtained using the usual method of producing CGH (b) versus using incomplete CGH with filled the interspersed holograms (d). Columns (a) and (c) are the corresponding CGHs that was programmed into the DMD.}
\label{fig:superposition}
\end{figure}
  
Different superposed beams obtained with eqns. \ref{eqn:combined} and \ref{eqn:usual} are presented in fig. \ref{fig:superposition}. Figure \ref{fig:superposition} (a) and (c) are the CGH encoded onto the DMD. As expected these are not the same. Figure \ref{fig:superposition} (b) and (d) beams are from these holograms. The generated beams obtained with  eqn. \ref{eqn:combined} resembles that obtained from eqn.\ref{eqn:usual} even at different beam combinations. We explain this by reviewing the concept of beam reconstruction with holograms. Upon reconstruction, the field right after the hologram is $E_{ref}H$ \cite{goodman2005intro}. For a single hologram $E_{ref}H\sim E_{ref}|E_{ref}+E_{beam}|^2$. This produces four terms: the two which are not diffracted, become the zeroth-order term while the other two become conjugate beams corresponding to the 1st order diffractions (+1 and -1, respectively). The 1st order diffractions are the reconstructed beams. In the case of beams with OAM, these beams have opposite topological charge signs but of the same values. With multiple beams in a hologram as in eqn. \ref{eqn:usual}, the reconstructed beams are proportional to the superposition of the $E$'s. It is therefore not surprising that eqn. \ref{eqn:usual} is used to generate superposed beams.

The situation is quite different when using eqn. \ref{eqn:combined} although as we will discuss later, the results are similar to the ones obtained in eqn. \ref{eqn:usual}. The field right after the hologram is $ E_{ref}H\sim E_{ref}(H_1^i+H_2^{ii}+...H_n^{i^n})$ where $H_n\sim |E_{ref}+E_n|^2$. Provided that the $E_{ref}$ are the same, our reconstructed beam will be $E_1^i+E_2^{ii}+...E_n^{i^n}$ and its conjugate propagating with the same angle but at the opposite side with respect to the zeroth order. The $E$'s are still in different domains.  This is where the concept of diffraction is necessary. The $E$'s as they propagate self-heals. In doing so, the domains of the $E$'s will overlap. Since the beam that impinges the different portions of the hologram comes from the same coherent source, the beam would now interfere producing a superposition, $E_1+E_2+...E_n$. Of course, the disadvantage of this method is that most of the energy of the beam used in the reconstruction will go to the zeroth order and hence, will produce superposed beams of lesser intensities as what we have observed in our experiments. 

The generation of superposed beams with different holograms interspersed together is \emph{not intuitive}. However, as we have  shown in our experiments and with our discussion, \emph{it is possible}. By combining our results in fig. \ref{fig:profiles}(d) with our new way of producing superposed beams, a superposed beam with different amplitudes, $a_1E_1+a_2E_2+...a_nE_n$can be generated. Our results here indicate that the OAM values are the same and that our initial numerical calculations point out that the mode components are also the same \cite{zambale2015incoherent}. Experimental mode component detection however, is beyond the scope of this study.

\section{Conclusion}
In summary, we are able to provide experimental evidence that incomplete CGHs can produce beams with OAM. We show and explain how the intensity of the reconstructed beam decreases with increasing information loss. We present proof that although the intensity is diminished, the topological charge is conserved. Our results also indicate that at least for beams with OAM, minute information loss is not a necessary parameter when one only needs to conserve the topological charge. The limit of information loss with which  the topological charge information will be diminished however, has not been attained due to the crudeness of detection. We believe that this is dependent on the size of the hologram matrix. In this end, mode projection can be done to determine this limit \cite{mair2001entanglement, steinlechner2016frequency}. Finally, we present a non-intuitive method to produce superposed beams for pixelated devices.  We believe our results are important especially since imperfections in programmable devices and in the fabrication of optical devices are sometimes unavoidable. 

\bigskip
The authors acknowledge the Office of the Chancellor of the University of the Philippines Diliman, through the Office of the Vice Chancellor for Research and Development, for funding support through the Outright Research Grant. N. Hermosa is a recipient of the Balik PhD Program of the Office of the Vice President for Academic Affairs of the University of the Philippines (OVPAA-BPhD-2015-06). NA Zambale is a scholar of the Department of Science and Technology-Science Education Institute (DOST-SEI) of the Republic of the Philippines. 

\bibliographystyle{model1-num-names}
\bibliography{sample.bib}

\end{document}